\input phyzzx
\hsize=417pt 
\sequentialequations
\Pubnum={ EDO-EP-9 }
\date={ \hfill March 1997 }
\titlepage
\vskip 32pt
\title{ Evaporation of Three Dimensional Black Hole in Quantum 
Gravity }
\author{Ichiro Oda \footnote\dag {E-mail address: 
sjk13904@mgw.shijokyo.or.jp}}
\vskip 12pt
\address{ Edogawa University,                                
          474 Komaki, Nagareyama City,                        
          Chiba 270-01, JAPAN     }                          
%
%
%
%
%
\abstract{ We discuss an evaporation of (2+1)-dimensional 
black hole by using quantum gravity holding in the vicinity of the black 
hole horizon. It is shown that the black hole evaporates at a definite 
rate by emitting matters through the quantum tunneling effect. A relation 
of the present formalism to the black hole entropy is briefly commented. 
 } 
\endpage
%
%
%

\def\sp(#1){\noalign{\vskip #1pt}}

%
%
%
%
%
\topskip 30pt
\par
The canonical formalism provides us with a useful starting point for 
quantization by Dirac [1]. In particular when applied for general 
relativity, the dynamics is entirely controlled by constraints [1, 2]. It 
is well known that their imposition as operator equations on the physical 
states produces the Wheeler-DeWitt equation [3] although its physical 
interpretation is sometimes unclear and difficult. Our aim in the present 
study is to make use of the canonical quantization formalism for a system 
having a black hole in order to understand the black hole evaporation [4] 
in three dimensions in the framework of quantum gravity. 

Since an announcement by Hawking [4] that black holes are not, after all, 
completely black, but emit a thermal radiation due to quantum effects, 
there have been many efforts to extend his semiclassical analysis where 
the background metric is treated classically, but the 
matter fields quantum mechanically, to a completely quantum-mechanical 
analysis. At the moment, it seems to be fair to say that we do not yet 
have a fully satisfactory and consistent theory of quantum gravity. 
Recently, however, there appeared one interesting and fully
quantum-mechanical approach that is based on 
both the black-hole minisuperspace model and 
the canonical quantization formalism, and was applied for understanding
of the black hole radiation [5, 6] and the mass inflation 
in four dimensions [7]$\footnote\dag
{P.Moniz has independently considered a similar model from a different 
motivation (private communication).}$. The key 
observation is that near the black hole horizon the 
Hamiltonian constraint becomes proportional to the supermomentum 
constraint (we will later clarify the reason) and, at the same time, has 
a tractable form by which we can solve the Wheeler-DeWitt equation
analytically. 

Even if not completely satisfactory, it seems to be very reasonable
at least for the present author to consider a theory of quantum gravity 
in the vicinity of the black hole horizon by the 
following reasons: it is nowadays thought that some important 
properties of quantum black holes such as the black hole thermodynamics 
[8] and the Hawking radiation [4] have an origin of the existence of the 
horizon [9, 10, 11]. More recently, Carlip has remarkably derived 
the entropy of the three dimensional black hole by counting 
the microscopic states associated with the horizon [12]. 

In this article, by means of an extended formalism of the above-mentioned 
formalism [5-7] we would like to consider an evaporation of 
(2+1)-dimensional black hole which was recently discovered by Ba$\tilde 
n$ados et al. [13] since one expects that questions about quantum black 
holes can be explored in considerable detail without the unnecessary 
complications coming from higher spacetime dimensions. Moreover, it 
is observed that in the Lovelock gravity the two branches of black holes 
emerge, namely, one for even dimensions, with strong similarities to the 
Schwarzschild black hole in four dimensions, and another for odd 
dimensions, with common features with three dimensional black hole [14].
Thus it is of interest to study whether the previous formalism or its 
extended version is applicable to the present case and leads to a 
physically meaningful result or not.

Let us start by constructing a canonical formalism of a spherically symmetric 
system with a black hole in three dimensions. An analogous formalism in 
four dimensions has already constructed in the Refs.[6, 15]. 

The three dimensional action that we consider is of the form
$$ \eqalign{ \sp(2.0)
S = \int \ d^3 x \sqrt{-^{(3)}g} \bigl[ {1 \over {16 \pi G}} \bigl( 
{}^{(3)}R + {2 \over l^2} \bigr) - {1 \over 4 \pi} {}^{(3)}g^{\mu\nu} 
(D_{\mu} \Phi)^{\dag} D_{\nu} \Phi -  {1 \over 
16 \pi } F_{\mu\nu}F^{\mu\nu} \bigr],
\cr
\sp(3.0)} \eqno(1)$$
where $l$ is the scale parameter with dimension of length and is related 
to the cosmological constant by $l = {1 \over \sqrt{- \Lambda}}$, 
$\Phi$ is a complex scalar field with the electric charge $e$,  
$$ \eqalign{ \sp(2.0)
D_{\mu} \Phi = \partial_{\mu}\Phi + i e A_{\mu} \Phi
\cr
\sp(3.0)} \eqno(2)$$
is its covariant derivative, $A_{\mu}$ is the electromagnetic field, 
and $F_{\mu\nu}$ is the corresponding field strength as usual given by
$$ \eqalign{ \sp(2.0)
F_{\mu\nu} = \partial_{\mu} A_{\nu} - \partial_{\nu} A_{\mu}.
\cr
\sp(3.0)} \eqno(3)$$
To clarify the three dimensional meaning we put the suffix $(3)$ 
in front of the metric tensor and the curvature scalar. 
We will follow the conventions adopted in the MTW textbook 
[16] and use the natural units $G = \hbar = c = 1$. The Greek indices 
$\mu, \nu, ...$ take 0, 1 and 2, and the Latin indices 
$a, b, ...$ take 0 and 1. Of course, the inclusion of other matter 
fields and the surface term in this formalism is straightforward even 
if we limit ourselves to the action (1) for simplicity. 

The most general spherically symmetric assumption for the metric is
$$ \eqalign{ \sp(2.0)
ds^2 &= {}^{(3)}g_{\mu\nu} dx^{\mu} dx^{\nu},
\cr
     &= g_{ab} dx^a dx^b + \phi^2 d\theta^2, 
\cr
\sp(3.0)} \eqno(4)$$
where the two dimensional metric $g_{ab}$ and the radial function $\phi$ 
are the function of only the two dimensional coordinates $x^a$. 
And the angular variable $\theta$ takes the value from 0 to $2 \pi$.
For the charged matter and the electromagnetic potential we take the
spherical ansatz $D_{\theta} \Phi = A_{\theta} =0$.
The substitution of these ansatz into (1) and then integration over the angular 
variable $\theta$ leads to the following effective action in two 
dimensions:
$$ \eqalign { \sp(2.0)
S &= {1 \over 8} \int \ d^2 x \sqrt{-g} \  \phi \ \bigl( R + {2 \over l^2} 
\bigr) - {1 \over 2} \int \ d^2 x \sqrt{-g} \ \phi \  g^{ab} (D_a 
\Phi)^{\dag} D_b \Phi 
\cr
&\qquad- {1 \over 8} \int \ d^2 x \sqrt{-g} \ \phi \  F_{ab} 
F^{ab}.
\cr
\sp(3.0)} \eqno(5)$$
Here let us make a brief comment on a curious feature of the 
gravitational sector in this action. 
It is easy to show that the dimensional reduction of the Einstein-Hilbert 
action with the cosmological constant from higher dimensions 
to two dimensions under the 
spherically symmetric ansatz in general leads to 
$$ \eqalign { \sp(2.0)
S = {1 \over 8} \int \ d^2 x \sqrt{-g} \  \bigl[ Y(\phi) R + V(\phi) + 
Z(\phi) g^{\mu\nu} \partial_{\mu} \phi \partial_{\nu} \phi \bigr], 
\cr
\sp(3.0)} \eqno(6)$$
which is a general class of the dilaton gravity. The model in hand, which 
is obtained by the reduction from three to two dimensions, has an 
interesting feature $Z(\phi) = 0$ and $Y(\phi) \propto V(\phi) \propto 
\phi$ which was previously introduced by 
Jackiew and Teitelboim [17] as a model of gravity in two dimensions 
since the Einstein-Hilbert action is a surface term in two dimensions. 
Subsequently, this de Sitter-type action was rewritten in terms of the 
topological BF theory [18]
$$ \eqalign { \sp(2.0)
S =  \int_M \ Tr B \ F. 
\cr
\sp(3.0)} \eqno(7)$$
Since it is known that the topological BF theory is exactly solvable 
[19-21], 
this interesting structure of the present model (5) would be of benefit 
to, for instance, an understanding of the black hole entropy in the 
future from the following observation. Recall that the geometry of the 
Euclidean black hole is largely independent of the spacetime dimension 
except the difference of even and odd dimensions, 
and then the black hole in $d$ dimensions has in general the topology 
of $R^2 \times 
S^{d-2}$, and the entropy comes quite generically from the existence of a 
possible conical singularity in the $R^2$ plane [22]. One therefore 
expects that a detailed analysis of the effective two dimensional 
theories such as (5) has a possibility of bringing us important 
informations in the black hole thermodynamics.

Now let us rewrite the action (5) into the first-order ADM form [2]. To do 
so, we shall foliate the region outside the black hole horizon by $x^0 = 
const$ spacelike hypersurfaces. 
The appropriate ADM splitting of (1+1)-dimensional spacetime is given by 
$$ \eqalign{ \sp(2.0)
g_{ab} = \left(\matrix{ { - \alpha^2 + {\beta^2 \over \gamma}} & \beta \cr
              \beta & \gamma \cr} \right),
\cr
\sp(3.0)} \eqno(8)$$
and the normal unit vector orthogonal to the hypersurface 
$x^0 = const$ becomes 
$$ \eqalign{ \sp(2.0)
n^a = ( {1 \over \alpha}, - {\beta \over {\alpha \gamma}}).
\cr
\sp(3.0)} \eqno(9)$$
Following the analogous procedure in the Refs.[6, 15], the action (5) can 
be written as
$$ \eqalign{ \sp(2.0)
S &= \int d^2x \  L  = \int d^2x \ \bigl[ \ {1 \over 4} \alpha \sqrt{\gamma} 
\bigl( - K n^a \partial_a \phi + {\alpha^\prime \over {\alpha\gamma}} 
\phi^\prime  + {1 \over l^2} \phi \bigr)  
\cr
&\qquad+ {1 \over 2} \alpha \sqrt{\gamma} \phi \bigl( | n^a D_a \Phi |^2
   - {1 \over \gamma} |D_1 \Phi |^2  \bigr) + 
   {1 \over 4} \alpha \sqrt{\gamma} \phi E^2 \bigr], 
\cr
\sp(3.0)} \eqno(10)$$
where
$$ \eqalign{ \sp(2.0)
E&= {1 \over \sqrt{-g}} F_{01} = {1 \over \alpha \sqrt{\gamma}}    
(\dot A_1 - A_0^{\prime}),
\cr
\sp(3.0)} \eqno(11)$$
and the trace of the extrinsic curvature $K = g^{ab} K_{ab}$ is expressed 
by 
$$ \eqalign{ \sp(2.0)
K = {\dot \gamma \over {2\alpha\gamma}} - {\beta^\prime \over 
{\alpha\gamma}} + {\beta \over {2\alpha\gamma^2}} \gamma^\prime.
\cr
\sp(3.0)} \eqno(12)$$
Here ${\partial \over {\partial x^0}} = \partial_0$ and ${\partial \over 
{\partial x^1}} = \partial_1$ are also denoted by an overdot and a prime, 
respectively. 

By differentiating the action (10) with respect to the canonical 
variables $\Phi (\Phi^{\dag}), \phi, \gamma$ and $A_1$, we have the 
corresponding conjugate momenta $p_{\Phi} ({p_{\Phi^{\dag}}}), p_{\phi}, 
p_{\gamma}$ and $p_A$ 
$$ \eqalign{ \sp(2.0)
p_{\Phi} =  {\sqrt{\gamma} \over 2} \phi (n^a D_a\Phi)^{\dag},
\cr
\sp(3.0)} \eqno(13)$$
$$ \eqalign{ \sp(2.0)
p_{\phi} = - {\sqrt{\gamma} \over 4}  K,
\cr
\sp(3.0)} \eqno(14)$$
$$ \eqalign{ \sp(2.0)
p_{\gamma} = - {1 \over 8 \sqrt \gamma} n^a \partial_a \phi,
\cr
\sp(3.0)} \eqno(15)$$
$$ \eqalign{ \sp(2.0)
p_A = {1 \over 2} \phi E.
\cr
\sp(3.0)} \eqno(16)$$
Then the Hamiltonian $H$ which is defined as
$$ \eqalign{ \sp(2.0)
H = \int dx^1 ( p_{\Phi} \dot \Phi + {p_{\Phi^{\dag}}} 
\dot \Phi^{\dag} 
+ p_{\phi} \dot \phi + p_{\gamma} \dot \gamma + p_A \dot A_1 - L ) 
\cr
\sp(3.0)} \eqno(17)$$
is expressed by a linear combination of constraints as expected 
$$ \eqalign{ \sp(2.0)
H = \int dx^1 ( \alpha H_0 + \beta H_1 + A_0 H_2 ), 
\cr
\sp(3.0)} \eqno(18)$$
where
$$ \eqalign{ \sp(2.0)
H_0 &= {2 \over {\sqrt{\gamma} \phi}} p_{\Phi} p_{\Phi^{\dag}} 
- 8 \sqrt{\gamma} p_{\phi} p_{\gamma}
+{1 \over 4} ({\phi^\prime \over {\sqrt{\gamma}}})^\prime
- {\sqrt{\gamma} \over 4} {1 \over l^2} \phi 
\cr
&\qquad+ {\phi \over {2 \sqrt{\gamma}}} | D_1 \Phi |^2  
+ {\sqrt{\gamma} \over \phi} p_A ^2 , 
\cr
\sp(3.0)} \eqno(19)$$
$$ \eqalign{ \sp(2.0)
H_1 = {1 \over \gamma}  [p_\Phi D_1 \Phi + {p_{\Phi^{\dag}}} 
(D_1 \Phi)^{\dag}]
 + {1 \over \gamma} p_\phi \phi^\prime - 2 p_\gamma ^\prime - {1 \over 
\gamma} p_\gamma \gamma^\prime, 
\cr
\sp(3.0)} \eqno(20)$$
$$ \eqalign{ \sp(2.0)
H_2 = - ie (p_{\Phi} \Phi - p_{\Phi^{\dag}} \Phi^{\dag}) -  p_A ^\prime.
\cr
\sp(3.0)} \eqno(21)$$
Note that $\alpha$, $\beta$ and $A_0$ are non-dynamical Lagrange 
multiplier fields.

The action can be cast into the first-order ADM canonical form by the 
dual Legendre transformation 
$$ \eqalign{ \sp(2.0)
S = \int d x^0 \bigl[ \int d x^1 ( p_{\Phi} \dot \Phi + 
{p_{\Phi^{\dag}}} \dot \Phi^{\dag} 
+ p_{\phi} \dot \phi + p_{\gamma} \dot \gamma + p_A 
\dot A_1 ) - H \bigr]. 
\cr
\sp(3.0)} \eqno(22)$$
As Regge and Teitelboim pointed out [23], in order to have the correct 
Hamiltonian which produces the Einstein equations through the Hamilton 
equations, one has to supplement the surface term to the Hamiltonian 
(22). It is straightforward to show that if one takes the boundary 
condition such that a contribution to the surface term from the apparent 
horizon vanishes the surface term from the spatial infinity produces 
the ADM mass of the black hole.

We now turn our attention to an application of the canonical formalism 
constructed in the above for understanding of an evaporation of the 
(2+1)-dimensional black hole [13] from the viewpoint of quantum gravity. 
To consider the simplest model of the black hole radiation, let us turn 
off the electromagnetic field and deal with the neutral scalar field by
which the constraint $H_2$ generating the $U(1)$ gauge transformations
identically vanishes. Moreover, we shall use an ingoing Vaidya metric 
to express the black hole radiation. The treatment of the case of the 
outgoing Vaidya metric can be made in a perfectly similar way. 

First of all, let us define the two dimensional coordinate $x^a$ by 
$$ \eqalign{ \sp(2.0)
x^a = (x^0, x^1) = (v - r, r),
\cr
\sp(3.0)} \eqno(23)$$
where the advanced time coordinate is defined as $v = t + r^*$ with the 
tortoise coordinate $dr^* = {dr \over {-g_{00}}}$. Then we shall fix the 
gauge freedoms corresponding to the two dimensional 
reparametrization invariances by the gauge conditions
$$ \eqalign{ \sp(2.0)
g_{ab} &= \left(\matrix{ { - \alpha^2 + {\beta^2 \over \gamma}} & \beta \cr
              \beta & \gamma \cr} \right),
\cr
 &= \left(\matrix{ - ( - M + {r^2 \over l^2})   &  1 + M - {r^2 \over l^2}
               \cr
              1 + M - {r^2 \over l^2} & 2 + M - {r^2 \over l^2} \cr} 
              \right),
\cr
\sp(3.0)} \eqno(24)$$
where the scale parameter $l$ and the black hole mass $M$ are now generally 
the function of the two dimensional coordinates $x^a$. Note that we have 
not fixed the gauge symmetries completely for later convenience though we 
may usually set the scale function $l$ to be constant by means of the 
remaining one gauge freedom. At present this does not cause any trouble 
since , afterward, we effectively fix this gauge symmetry in making an 
assumption of dynamical fields near the horizon.

{}From these equations the two dimensional line element takes a form of the 
Vaidya metric corresponding to the three dimensional black hole without 
rotation 
$$ \eqalign{ \sp(2.0)
ds^2 &= g_{ab} dx^a dx^b,
\cr
     &= - ( - M + {r^2 \over l^2}) dv^2 + 2 dv dr.
\cr
\sp(3.0)} \eqno(25)$$
Since we would like to study a dynamical black hole, it is useful to consider 
the local definition of horizon, i.e., the apparent horizon, instead of 
the global one, the event horizon. The apparent horizon is 
then defined as
$$ \eqalign{ \sp(2.0)
r_{+} = l \ \sqrt M.
\cr
\sp(3.0)} \eqno(26)$$

Since we have constructed the canonical formalism of a spherically 
symmetric system with a black hole in three dimensions, let us perform a 
canonical quantization. Following Dirac [1], we have to impose the constraints 
on the states as the operator equations and solve them. However, even in 
a rather simple setting of the present model, it is quite difficult to 
solve the constraints' equations owing to their complicated form. 
We therefore need some approximation method that 
retains the important features of the black hole physics. 
One of such approximations was proposed by Tomimatsu [5]. 
His critical idea is to solve the Hamiltonian and supermomentum 
constraints only in the vicinity of the apparent horizon. We will see 
that this approximation scheme yields to a remarkable simplification 
in the model at hand.
 
Near the apparent horizon, we shall make an assumption 
$$ \eqalign{ \sp(2.0)
\Phi \approx \Phi(v), \phi \approx r, l \approx l(v), M \approx const.
\cr
\sp(3.0)} \eqno(27)$$
We shall use $\approx$ to indicate the equalities which hold 
approximately near the apparent horizon from now on. In the latter two 
equations in (27), we have made a specific assumption in three 
dimensions. The reason is that in three dimensions or even in general odd 
dimensions the mass function $M$ 
is dimensionless so that it is not the mass but the scale function $l$ 
with dimension of length that plays a key role in describing the 
black hole properties [13, 14]. Indeed it is valuable to note that (27) 
is equivalent to an assumption where the radius of the black hole is in 
itself a dynamical function as seen later explicitly. Incidentally
 one can prove the above 
assumption (27) to be consistent with the field equations near the 
apparent horizon in an analogous way to our previous work [6].

Eq.(24) yields near the apparent horizon (26) 
$$ \eqalign{ \sp(2.0)
\alpha \approx {1 \over \sqrt 2}, \beta \approx 1, \gamma = {1 \over 
\alpha^2} \approx 2,
\cr
\sp(3.0)} \eqno(28)$$
and the canonical conjugate momenta (13)-(15) are given 
approximately as
$$ \eqalign{ \sp(2.0)
p_{\Phi} &\approx {1 \over 2} \phi \partial_v \Phi,
\cr
p_{\phi} &\approx  {1 \over 8} {M \over l} \partial_v l - {3 \over 8} 
{\sqrt M \over l},
\cr
p_{\gamma} &\approx {1 \over 16}.
\cr
\sp(3.0)} \eqno(29)$$
Moreover, after a careful calculation the two constraints are 
proportional to each other
$$ \eqalign{ \sp(2.0)
{1 \over \sqrt 2} H_0 &\approx H_1,
\cr
             &\approx {2 \over \phi} p_{\Phi}^2 -  p_\phi - {3 \over 8} 
             {\sqrt M \over l}. 
\cr
\sp(3.0)} \eqno(30)$$

At this stage we would like to think of the reason why the Hamiltonian and the 
supermomentum constraints have become  proportional to each other near 
the horizon since the previous works [5-7] are obscure in this respect.
  The Hamiltonian constraint $H_0$ and supermomentum constraint 
$H_1$ generate the time translation and the spatial 
displacement, respectively, given by
$$ \eqalign{ \sp(2.0)
H_0 : x^0 &\rightarrow x^0 + \varepsilon^0, 
\cr
H_1 : x^1 &\rightarrow x^1 + \varepsilon^1,           
\cr
\sp(3.0)} \eqno(31)$$
where $\varepsilon^a$ is the infinitesimal transformation parameter. From 
(23), it is shown that
$$ \eqalign{ \sp(2.0)
H_0 : v &\rightarrow v + \varepsilon^0 + \varepsilon^1, 
\cr
H_1 : r &\rightarrow r + \varepsilon^1.           
\cr
\sp(3.0)} \eqno(32)$$
Note that the time is standing still at the apparent horizon, so that the 
time translation associated with the transformation parameter 
$\varepsilon^0$ is frozen there. Thus in the coordinates $(v, r)$ the 
only nontrivial gauge motion is nothing but the spatial displacement at 
the horizon. This observation is also certified by the equalities from 
(18), (28) and (30) that $\alpha H_0 \approx \beta H_1 \approx H_1$.

By imposing the constraint (30) as an operator equation on the state, 
one obtains the Wheeler-DeWitt equation
$$ \eqalign{ \sp(2.0)
i {\partial \Psi \over {\partial T}} = \bigl( - {\partial^2 \over 
{\partial \Phi^2}} - {3 \over 16} M \bigr) \Psi,
\cr
\sp(3.0)} \eqno(33)$$
where we have introduced $T = \log{({1 \over \phi})^2}$. 
This Wheeler-DeWitt equation can be interpreted as the Schr$\ddot o$dinger 
equation with the time $T$ and the Hamiltonian $H = p_{\Phi}^2 - {3 \over 
16} M$ in the superspace. 

Now it is easy to find a special solution of the above Wheeler-DeWitt 
equation by the method of separation of variables. The result is 
$$ \eqalign{ \sp(2.0)
\Psi = (B e^{i \sqrt A \Phi(v)} + C e^{-i \sqrt A \Phi(v)} ) \ e^{- i( A 
- {3 \over 16} M ) T},
\cr
\sp(3.0)} \eqno(34)$$
where $A$, $B$, and $C$ are integration constants. If one defines an 
expectation value $< \cal O >$ of an operator $\cal O$ in a rather naively 
as
$$ \eqalign{ \sp(2.0)
< {\cal O} > = {1 \over {\int d\Phi |\Psi|^2 }} \int d\Phi \Psi^* {\cal O} 
\Psi,
\cr
\sp(3.0)} \eqno(35)$$
one can calculate a change rate of the radius of a black hole horizon 
$< \partial_v r_{+} >$ through the absorption  or the emission of 
the neutral scalar matters by using (26) and either (29) or (30)
$$ \eqalign{ \sp(2.0)
< \partial_v r_{+} > =  {16 A \over M}.
\cr
\sp(3.0)} \eqno(36)$$
This equation shows the absorption of the external matters by a black 
hole when one chooses the 
constant $A$ to be a positive constant, e.g., ${1 \over 16} k_1 ^2$.  
Then the change rate of the radius of a black hole horizon becomes
$$ \eqalign{ \sp(2.0)
< \partial_v r_{+} > = {k_1 ^2 \over M} ,
\cr
\sp(3.0)} \eqno(37)$$
and the physical state is given by
$$ \eqalign{ \sp(2.0)
\Psi = (B e^{i {1 \over 4} | k_1 | \Phi(v)} + C e^{-i {1 \over 4} | k_1 | 
\Phi(v)} )  \ e^{- i {1 \over 16} (k_1 ^2 - 3 M) T}.
\cr
\sp(3.0)} \eqno(38)$$
On the other hand, if one takes the constant $A$ to be a negative 
constant, e.g., $- {1 \over 16} k_2 ^2$, (36) means the Hawking 
radiation: 
$$ \eqalign{ \sp(2.0)
< \partial_v r_{+} > = - {k_2 ^2 \over M} ,
\cr
\sp(3.0)} \eqno(39)$$
with the physical state
$$ \eqalign{ \sp(2.0)
\Psi = (B e^{ - {1 \over 4} | k_2 | \Phi(v)} + C e^{ {1 \over 4} | k_2 | 
\Phi(v)} )  \ e^{ i {1 \over 16} (k_2 ^2 + 3 M) T}.
\cr
\sp(3.0)} \eqno(40)$$
Note that if one chooses the boundary condition $C=0$ in (38) and (40), 
the physical state (38) represents the scalar wave propagating in black 
hole from the exterior region across the horizon while the state (40) is 
exponentially damping tunneling state in classically forbidden region. 
These behavior of the physical states seems to be physically plausible 
with our interpretation that (37) describes the absorption of matters by 
a black hole, on the other hand, (39) represents the the Hawking 
radiation through the quantum tunneling effect. 
Incidentally, it is of interest to comment here that if, instead of (27), 
we adopt an alternative assumption $\Phi \approx \Phi(v), \phi \approx 
r, M \approx M(v)$, and $l \approx const.$ as done in the case of the four 
dimensional Schwarzschild black hole (precisely speaking, in four 
dimensions, $\Lambda = - {1 \over l^2} = 0$), it is shown that we obtain 
the same result as (36). This correspondence stems from the mathematical 
fact that both assumptions share the common Wheeler-DeWitt equation. 
In this sense, our result is independent of these assumptions.

Now let us consider the physical meaning of the result in the case of the 
Hawking radiation (39) and (40). Eq.(39) implies that the radius of the 
black hole horizon gradually descreases at a definite rate by emitting 
the thermal radiation, and then evaporates completely. However, it is 
dubious to extrapolate this result literally to the endpoint of 
evaporation since the physical state (40) fluctuates strongly there 
owing to 
the huge $T$ factor. This huge quantum fluctuation might be related 
to the quantum instability of the black hole singularity in three 
dimensions [24].

To summarize, in this article we have analysed the black hole radiation 
in three spacetime dimensions by using the quantum gravity holding in the 
vicinity of the black hole horizon. We have seen that the black hole 
radius descreases gradually by emitting the thermal radiation. Here
it is worth pointing out an advantage of our formalism compared to 
the Hawking's original formulation.  In the Hawking's 
semiclassical approach the gravitational field is fixed as a classical 
background and only the matter field is treated to be quantum-mechanical. 
By contrast, our present formulation is purely quantum-mechanical.
We hope that the present formalism may have some implications in the
membrane paradigm of quantum black holes in the future.

\vskip 32pt
\leftline{\bf References}
\centerline{ } %
\par
\item{[1]} P.A.M.Dirac, Lectures on Quantum Mechanics (Yeshiva University, 
1964). 

\item{[2]} P.Arnowitt, S.Deser, and C.W.Misner, in Gravitation: An 
Introduction to Current Research, edited by L.Witten (Wiley, New York, 
1962).

\item{[3]} B.S.DeWitt, Phys. Rev. {\bf 160} (1967) 1113. 

\item{[4]} S.W.Hawking, Comm. Math. Phys. {\bf 43} (1975) 199.

\item{[5]} A.Tomimatsu, Phys. Lett. {\bf B289} (1992) 283.

\item{[6]} A.Hosoya and I.Oda, ``Black Hole Radiation inside Apparent 
Horizon in Quantum Gravity'', Prog. Theor. Phys. (in press).

\item{[7]} I.Oda, ``Mass Inflation in Quantum Gravity'', EDO-EP-8, 
gr-qc/9701058.

\item{[8]} J.D.Bekenstein, Phys. Rev. {\bf D7} (1973) 2333.

\item{[9]} K.S.Thorne, R.H.Price and D.A.Macdonald, Black Hole: The 
Membrane Paradigm (Yale University Press, 1986).
 
\item{[10]} G.'t Hooft, Nucl. Phys. {\bf B335} (1990) 138; Phys. Scripta 
{\bf T15} (1987) 143; ibid. {\bf T36} (1991) 247.

\item{[11]} I.Oda, Int. J. Mod. Phys. {\bf D1} (1992) 355; Phys. Lett. 
{\bf B338} (1994) 165; Mod. Phys. Lett. {\bf A10} (1995) 2775.

\item{[12]} S.Carlip, Phys. Rev. {\bf D51} (1995) 632; ``The Statistical 
Mechanics of the Three-Dimensional Euclidean Black Hole'', UCD-96-13, 
gr-qc/9606043.

\item{[13]} M.Ba$\tilde n$ados, C.Teitelboim, and J.Zanelli, Phys. Rev. 
Lett. {\bf 69} (1992) 1849; M.Ba$\tilde n$ados, M.Henneaux, C.Teitelboim, 
and J.Zanelli, Phys. Rev. {\bf D48} (1993) 1506.

\item{[14]} M.Ba$\tilde n$ados, C.Teitelboim, and J.Zanelli, Phys. Rev. 
{\bf D49} (1994) 975.

\item{[15]} P.Thomi, B.Isaak, and P.Hajicek, Phys. Rev. {\bf D30} (1984) 
1168; P.Hajicek, Phys. Rev. {\bf D30} (1984) 1178.

\item{[16]} C.W.Misner, K.S.Thorne, and J.A.Wheeler, Gravitation (Freeman,
1973). 

\item{[17]} R.Jackiew, in Quantum Theory of Gravity, edited by 
S.Christensen (Adam Hilger, Bristol, 1984); ibid. C.Teitelboim.

\item{[18]} T.Fukuyama and K.Kamimura, Phys. Lett. {\bf B160} (1985) 259; 
K.Isler and C.Trugenberger, Phys. Rev. Lett. {\bf 63} (1989) 834; 
A.Chamseddine and D.Wyler, Phys. Lett. {\bf B228} (1989) 75.

\item{[19]} G.T.Horowitz, Comm. Math. Phys. {\bf 125} (1989) 
417.

\item{[20]} I.Oda and S.Yahikozawa, Phys. Lett. {\bf B234} (1990) 69; 
ibid. {\bf B238} (1990) 272; Class. Quant. Grav.{\bf 11} (1994) 2653.

\item{[21]} M.Blau and G.Thompson, Ann. Phys. (N.Y.) {\bf 205} (1991) 130.

\item{[22]} G.W.Gibbons and S.W.Hawking, Phys. Rev. {\bf D15} (1977) 
2753; S.Carlip and C.Teitelboim, Phys. Rev. {\bf D51} (1995) 622.

\item{[23]} T.Regge and C.Teitelboim, Ann. Phys. (N.Y.) {\bf 88}
 (1974) 286.

\item{[24]} I.Oda, ``Quantum Instability of Black Hole Singularity in 
Three Dimensions'', EDO-EP-10.

\endpage
%

%
\bye